\documentclass{article}
\RequirePackage[OT1]{fontenc}
\RequirePackage{amsthm,amsmath,amsfonts}
\RequirePackage[authoryear]{natbib}
\RequirePackage[colorlinks,citecolor=blue,urlcolor=blue]{hyperref}
\usepackage[a4paper, total={6.5in, 8.5in}]{geometry}
\usepackage{latexsym,graphicx,subcaption,verbatim,nameref}
\usepackage{diagbox}
\usepackage[affil-it]{authblk}
\usepackage[all,cmtip]{xy}
\usepackage[toc,title,titletoc]{appendix}
\usepackage[capposition=top]{floatrow}
\usepackage{rotating}
\usepackage{threeparttable,booktabs}
\usepackage{csquotes}
\usepackage{chngcntr}
\usepackage{subcaption}
\usepackage{graphicx}
\usepackage{esint}
\usepackage{relsize}
\usepackage{color}
\usepackage{multirow}
\usepackage{algorithm}
\usepackage{algpseudocode}
\counterwithin{figure}{section}
\counterwithin{table}{section}
\numberwithin{equation}{section}
\theoremstyle{plain}
\newtheorem{theorem}{Theorem}[section]

\theoremstyle{definition}

\newtheorem{remark}[theorem]{Remark}
\newtheorem{example}[theorem]{Example}

\newcommand*{\Scale}[2][4]{\scalebox{#1}{$#2$}}%

\providecommand{\keywords}[1]{\textbf{\textit{Keywords---}} #1}
\let\oldref\ref
\renewcommand{\ref}[1]{(\oldref{#1})}

\begin{document}

\title{A novel approach to estimate the Cox model with temporal covariates and application to medical cost data}

\author[1]{Xiaoqi Zhang%
\thanks{Email: \texttt{xiaoqizh@buffalo.edu}}}

\author[2]{Xiaobing Zhao%
\thanks{Email: \texttt{maxbzhao@126.com}}}

\author[1]{Yanqiao Zheng%
\thanks{Email: \texttt{yanqiaoz@buffalo.edu}; Corresponding Author. Address: No. 18, Xueyuan Street, Xiasha Higher Education Park, Hangzhou, Zhejiang, 310018, China}}
\affil[1]{School of Finance\\ Zhejiang University of Finance and Economics}
\affil[2]{School of Data Science\\ Zhejiang University of Finance and Economics}

\maketitle

\begin{abstract}
We propose a novel approach to estimate the Cox model with temporal covariates. Our new approach treats the temporal covariates as arising from a longitudinal process which is modeled jointly with the event time. Different from the literature, the longitudinal process in our model is specified as a bounded variational process and determined by a family of Initial Value Problems associated with an Ordinary Differential Equation. Our specification has the advantage that only the observation of the temporal covariates at the time to event and the time to event itself are required to fit the model, while it is fine but not necessary to have more longitudinal observations. This fact makes our approach very useful for many medical outcome datasets, like the New York State's Statewide Planning and Research Cooperative System (SPARCS) and the National Inpatient Sample (NIS), where it is important to find the hazard rate of being discharged given the accumulative cost but only the total cost at the discharge time is available due to the protection of patients' information. Our estimation procedure is based on maximizing the full information likelihood function. The resulting estimators are shown to be consistent and asymptotically normally distributed. Variable selection techniques, like Adaptive LASSO, can be easily modified and incorporated into our estimation procedure. The oracle property is verified for the resulting estimator of the regression coefficients. Simulations and a real example illustrate the practical utility of the proposed model. Finally, a couple of potential extensions of our approach are discussed.
\end{abstract}

\section{Introduction}\label{introduction}

In the proportional hazards model (\cite{cox1972regression,andersen1982cox}), the hazard function of the event time $T$ takes the form 

\begin{equation}\label{cox-proportional-hazard}
\lambda\left(t\mid Z\right)=\lambda_{0}\left(t\right)\exp\left(b_{0}^{T}Z\right)
\end{equation}
where $\lambda\left(t\mid Z\right)$ is the conditional hazard function
of $T$ given the $p\times1$ covariate vector $Z$, $\lambda_{0}\left(t\right)$
is an unspecified baseline hazard function and $b_{0}$ is a $p\times1$
vector of unknown regression coefficients. Although in the original
paper, the covariate $Z$ is viewed as random vectors independent
from time $t$, the model \eqref{cox-proportional-hazard} can be easily extended to the case where
$Z$ is time-dependent and $Z\left(t\right)$ is assumed to be given
through an unknown stochastic process. 

The main-stream procedure used to estimate the model \eqref{cox-proportional-hazard} is the maximum
partial likelihood (MPL) procedure, which applies well whether or not the covariates are time-dependent. However, when time-dependent covariates being involved, the consecutive observation of the covariates is required in the sense that every subject must have its covariates observed at all the failure time prior to its own failure. In the other words, let $T_i<T_j$ be the observed failure time for two different subjects $i$ and $j$, then for the dying later subject $j$, its covariate $Z_j$ must have values observed at both of $T_i$ and $T_j$. Otherwise, the MPL procedure won't work. Although some approximation methods were proposed to relax consecutive observation requirement as discussed in \cite{andersen1992repeated}, the MPL procedure can't be applied effectively to the medical cost datasets, like the New York State's Statewide Planning and Research Cooperative System (SPARCS), or the National Inpatient Sample (NIS), where it is important to find the hazard rate of being discharged given the accumulative cost while only the total cost for each inpatient observed at the discharge time is available. 
%
%

In this paper, we shall propose a novel estimation procedure for the
model \eqref{cox-proportional-hazard}, which can generate consistent estimates for parameters and the baseline hazard in the model \eqref{cox-proportional-hazard} even if only the observations  $\left\{\left(Z_{i,T_i},T_i\right):i=1,\dots,n\right\}$ are available, where all the observations $Z_{i,T_j}$ with $T_j<T_i$ are missing. Our procedure is based on joint modelling the longitudinal process that generates the time-dependent covariates and the time to event. The topic of joint model has been widely discussed (\cite{henderson2000joint,
song2002semiparametric,hsieh2006joint,
ye2008semiparametric,rizopoulos2011dynamic,
kim2013joint,lawrence2015joint}). A comprehensive review is also available in \cite{tsiatis2004joint, sousa2011review,ibrahim2010basic}. However, all these works require the condition that the number of observations of the longitudinal measurement before the time to event is greater than the dimension of the longitudinal process, which restricts their usefulness for the input data with only one observation of the covariate value at the event time.
%
In this paper, we propose an alternative specification of the joint model.
Formally, we only assume that the longitudinal measurements follow a bounded variational process that can be expressed as a stochastic integral as below:
%
\begin{equation}\label{potential-growth-process}
\mathcal{Z}(t):=Z_{0}+\int_{0}^{t}Y(s)\epsilon(s)ds.
\end{equation}
where $Z_{0}$ represents the initial value and $Y(t):=I\left(T>t\right)$. The model \eqref{potential-growth-process} consists of two components: 

(1) the longitudinal process:
\begin{equation}\label{longitudinal}
Z(t)=Z_0+\int_{0}^{t}\epsilon(s)ds
\end{equation}
which characterizes the evolution of the longitudinal measurements, we assume the conditional expectation of the increment rate $\epsilon(t)$ has the following parametric form
\begin{equation}\label{conditional-expectation-of-growth-rate}
q\left(z,t\mid a_{0}\right) := E\left(\epsilon(t)\mid Z(t)=z,a_{0}\right);
\end{equation} 
(2) the event process $Y$ which determines the time to event and its conditional expectation has the form \eqref{cox-proportional-hazard}, i.e.
\begin{equation}\label{conditonal-hazard}
\lambda\left(t\mid z\right)=
\lambda_{0}\left(t\right)\exp\left(b_{0}^{T}z\right):=
E\left(-dY(t)\mid Y(t^{-})=1, Z(t^{-})=z\right).
\end{equation} 
It turns out by \cite{XQZhang2017}, combining \eqref{conditional-expectation-of-growth-rate} and \eqref{conditonal-hazard} yields a complete specification of the joint model in the sense that if two joint models share a common pair of the conditional hazard function \eqref{conditonal-hazard} and the conditional expectation \eqref{conditional-expectation-of-growth-rate}, all the distributions in interest arising from the two joint models are identical. The equation \eqref{conditional-expectation-of-growth-rate} is the key to derive an explicit expression of the joint probability density function (pdf) of $Z_T$ and $T$ which helps design our estimation procedure. To our best
knowledge, there has not been any previous works attempting to specify the longitudinal process as
\eqref{conditional-expectation-of-growth-rate}. We hope our work could provide some hints to the future development of this field.

In model \eqref{conditional-expectation-of-growth-rate} and \eqref{conditonal-hazard}, there are three sets of parameters, $a_{0}$, $b_{0}$ and $\lambda_{0}$. Among them, $\lambda_{0}$ has infinite dimension. Our procedure will estimate the three kinds of parameters through maximizing the full information likelihood function, where the likelihood is constructed from the joint probability density function (pdf) of $Z_T$ and $T$. By \cite{XQZhang2017}, this joint pdf is expressed
as below by using the function $q$ and $\lambda$:

\begin{equation}\label{joint-pdf}
\Scale[0.8]{
pdf\left(z,t\mid a_{0},b_{0},\lambda_{0}\right)=\tilde{p}\left(z,t\right)\times\exp\left(-\int_{0}^{t}\lambda_{0}\left(t-s\right)\exp\left(b_{0}^{T}g\left(z,t,s\right)\right)ds
\right)\times\lambda_{0}\left(t\right)\exp
\left(b_{0}^{T}z\right)}
\end{equation}
where the function $\tilde{p}$ is the time-dependent pdf induced
by the longitudinal process $\left\{ Z_{t}\right\} $ and by \cite{XQZhang2017} it can be expressed as 
\begin{equation}\label{p-tilde}
\tilde{p}(z,t)=p\left(g\left(z,t,t\right),0\right)\cdot \mathcal{J}_{z,t},
\end{equation}
The function $p(,.0)$ is the initial pdf induced by $Z_0$ and for every $t$, $\mathcal{J}_{z}(t)$ denotes the Jacobian of the function $g\left(.,t,t\right)$ evaluated at the point $z$.  
The
function $g$ is solely determined by $q$ through solving a family
of initial value problems (IVPs). Namely for every fixed $z$ and $t$,
$g\left(z,t,.\right)$ is the solution to the following ordinary differential
equation (ODE) for $s\in\left(0,t\right)$:
\begin{equation}\label{ode}
z'\left(s\right) = -q\left(z\left(s\right),t-s\mid a_{0}\right)
\end{equation}
subject to the initial condition $g\left(z,t,0\right)=z$.

In addition to estimating model parameters, in practice it is also important to select the significant covariates. Variable selection approaches has been extensively studied by many authors. The least absolute shrinkage and selection operator (LASSO) was presented by \cite{tibshirani1996regression}. \cite{fan2001variable} developed the nonconcave penalized approach (SCAD) for variable selection which applies to likelihood-based estimation procedures, including the MPL procedure for the Cox model (\cite{fan2002variable}). \cite{zou2006adaptive} developed an Adaptive LASSO approach and showed its oracle property under a general set of conditions. The estimation procedure proposed in the current paper can be easily combined with those variable selection approaches. In particular, we will incorporate a modified version of the Adaptive LASSO into our procedure and verify its oracle property.

The rest of this paper is organized as follows. In Section 2, we will sketch the estimation procedures in detail.
The large sample properties of resulting estimators are stated in
Section 3. Simulation results and the application to real world data
are presented in Section 4. Section 5 discusses some extensions of
our model and concludes. All proofs are collected in Appendix.

\section{ESTIMATION PROCEDURE}

The estimation procedure is based on maximizing the full information likelihood
function which is formed as the product of joint pdf of the failure time $T\in[t,t+dt)$ and the observation
of the longitudinal measure $Z_{t}$ at time $t$. To deal with the
non-parametric $\lambda_{0}$, we adopt the method that approximates
$\lambda_{0}$ through a sequence of finite dimensional step-wise
functions, denoted as $\lambda_{n}^{s}$, with the number of steps
given by the sample size $n$. 

Throughout this section, we assume the data input for the estimation procedure only has the observation at the time to event, such as $\left\{\left(Z_{T_i},T_i\right):\,i=1,\dots,n\right\}$, while in the remark section, we will briefly discuss the adjustment of our procedure to deal with the case where more longitudinal observations are available.

\subsection{Likelihood Function}\label{likelihood function}

Define $A\subset\mathbb{R}^{d}$ as the domain of all possible values
of the parameter $a$, $a_{0}\in A$ as the true parameter. Similarly,
Define $B\subset\mathbb{R}^{p}$ as the domain of $b$, $b_{0}\in B$
as the true parameter. For every fixed $a\in A$, define $g\left(.\mid a\right)$
as the solution trajectories to the IVPs \eqref{ode} conditional on $a$. When the
analytic form of $g\left(.\mid a\right)$ is not available, we can
use its numerical approximation in place.
There are many efficient numerical solvers to the IVPs \eqref{ode}. In this
paper, we pick up the Euler's method \cite{} for the purpose of being simple and illustrative. Similarly, write $\mathcal{J}_{z|a}(t)$ as the Jacobian of $g\left(.,t,t\mid a\right)$ and when necessary it can be replaced by its numerical version.

By \cite{XQZhang2017}, the initial pdf $p(.,0)$ is uniquely determined by the function \eqref{potential-growth-process} and the joint pdf \eqref{joint-pdf}. In particularly, given the joint pdf \eqref{joint-pdf}, there is a well defined map $a\mapsto p_{a}(.,0)$ from the parameter space $A$ to the space of the pdfs over $\mathbb{R}^{p}$. Therefore, given the input data and a fixed parameter $a$, we can estimate $p_a(.,0)$ by the Gaussian kernel density method as below:
\begin{equation}\label{initial-appro}
p_{a,n}(z,0):=\frac{1}{n}\sum_{i=1}^{n}G_{n^{-1/4}}\left(z-g\left(z_{t_i},t_i,t_i\mid a\right)\right)
\end{equation}
where $G_h$ denote the Gaussian kernel function with kernel width $h$. In this paper, we simply select the kernel width as $\frac{1}{n^4}$ as can guarantee the function \eqref{initial-appro} converges to $p_{a}(.,0)$ in the $L^1$ norm for all $a$ \cite{}.

For the baseline hazard $\lambda$, without loss of generality,
we set $t_{0}=0$, $\theta_{0}=1$ and let $0\leq t_{1}<t_{2}<\cdots<t_{n}<\infty$
be the ordered statistics of the $n$ observed failure time. A step-wise
version of the non-parametric baseline hazard is constructed
as below:
\begin{equation}\label{step-wise-lambda}
\lambda_{n}^{s}\left(t\right):=\sum_{i=1}^{n}\theta_{i}\cdot I_{[t_{i-1},t_{i})}\left(t\right)
\end{equation}
where $\theta=\left\{ \theta_{i}\geq0:i=1,\dots,n\right\} $ is a
set of parameters to be estimated. For each profile of the parameters
$\Omega_{n}=\left(\lambda_{n}^{s},a,b\right)$, we can define the
log likelihood as below:
\begin{equation}\label{likelihood1}
l_n\left(a,b,\lambda_{n}^{s}\right):=\frac{1}{n}\sum_{i=1}^{n}\left(
\begin{aligned}
&\log p_{a,n}\left(g\left(z_{t_i},t_i,t_i\mid a\right),0\right)
+\log \mathcal{J}_{z_{t_i}|a}\left(t_i\right)+\log\lambda_{n}^{s}\left(t_{i}\right)\\
&+b^{T}z_{t_{i}}-\int_{0}^{t_{i}}\lambda_{n}^{s}\left(t_{i}-\tau\right)\exp\left(b^{T}g\left(z_{t_{i}},t_{i},\tau|a\right)\right)d\tau
\end{aligned}
\right).
\end{equation} 
The estimator resulting from maximizing \eqref{likelihood1} is denoted as $\hat{\lambda}^{s}$
and $\hat{b}$ and $\hat{a}$. 

\begin{remark} \label{remark1}
The first order condition of the optimization problem \eqref{likelihood1} indicates
the relation 
\begin{equation}\label{constraint-for-lambda}
\theta_{i}=\frac{1}{\sum_{j=i+1}^{n}\int_{0}^{t_{i}-t_{i-1}}\exp\left(b^{T}g\left(z_{t_{j}},
t_{j},t_{j}-t_{i}+\tau|a\right)\right)d\tau}
\end{equation}
 at the optimal point $\hat{\lambda}_{n}^{s}$ and $\hat{b}$ and
$\hat{a}$. Relation \eqref{constraint-for-lambda} can be inserted as a set of constraints back
into the optimization problem \eqref{likelihood1}, which helps sharply reduce the dimension
of the original problem. 
%
\end{remark}

\subsection{Variable Selection }\label{variable selection}

A penalty function can be naturally incorporated into the estimation and determine the non-zero component of the coefficients
$b$ automatically. There exist many types of penalty functions in the literature,
such as LASSO (\cite{tibshirani1996regression}), adaptive LASSO (\cite{zou2006adaptive}), SCAD (\cite{fan2001variable}), MCP. For convenience of calculation,
we choose the adaptive LASSO penalty function defined by:

\begin{equation}\label{penality-funcion}
P_{\Lambda}\left(b_{j}\right):=\Lambda\frac{\left|b_{j}\right|}{\left|\hat{b}_{j}\right|^{r}}
\end{equation}
where $b_{j}$ is the $j$-th component of the the vector $b$, $\hat{b}_{j}$
is a consistent estimator of $b_{j}$ (say the estimator generated by
maximizing Eq. \eqref{likelihood1}), $\Lambda$ is a tuning parameter and $r>0$ (for
simplicity of calculation, we let $r=2$ throughout the paper). 

The penalty function is added into the likelihood function \eqref{likelihood1} and forms the following maximization problem \eqref{likelihood1}:
\begin{equation}\label{likelihood4}
\max_{a,u,\lambda^{s}}\left(l_{n}\left(a,\hat{b}+u,\lambda^{s}\right)+\sum_{j=1}^{p}P_{\Lambda_{n}}
\left(\hat{b}_{j}+\frac{u}{\sqrt{n}}\right)\right)
\end{equation}
where the choice of tuning parameter $\Lambda_{n}$ depends on the
sample size $n$. Denote $\hat{a}_{AL}$, $\hat{\lambda}_{AL}^{s}$
and $\hat{b}_{AL}$ ($:=\hat{b}+\hat{u}_{AL}$ with $\hat{u}_{AL}$ being the solution to \eqref{likelihood4}) as the estimator corresponding
to maximize Eq. \eqref{likelihood4}. 

\begin{remark}
In the original work (\cite{zou2006adaptive}), a wide range of candidate values can be selected for the tuning parameter $\Lambda_n$ as long as the asymptotic property $\Lambda_n\cdot n^{-\frac{1}{2}}\rightarrow 0$ and $\Lambda_n\cdot n^{\frac{1}{2}}\rightarrow \infty$ hold. In practice, there are multiple ways to select the value of $\Lambda_n$ such as the Bayesian criterion method \cite{} which is based on iterative calculation of the values of the Bayesian criterion and the likelihood function. Instead of using the iterative procedure, we simply set $\Lambda_n=n^{-\frac{1}{4}}$  in order to reduce the computation load.  
\end{remark}

\begin{remark}
The original adaptive LASSO method is designed for the OLS procedure (\cite{zou2006adaptive}), but it turns out this method applies very well to the likelihood-based estimation procedures as ours. The oracle property of the adaptive LASSO estimator, $\hat{b}_{AL}$, will be proved in Appendix \ref{theorem-ALASSO}.
\end{remark}

\section{LARGE SAMPLE PROPERTIES}

The consistency and asymptotic normality of the estimators, $\hat{a}$,
$\hat{b}$ and $\hat{\lambda}_{n}^{s}$, constructed in section 2.1 will be established in this section. We will also show the oracle
property of the adaptive LASSO estimator $\hat{b}_{AL}$ and the consistency
and asymptotic normality of $\hat{\lambda}_{AL}^{s}$ and $\hat{a}_{AL}$.

\subsection{Large Sample Properties of $\hat{a}$, $\hat{b}$ and $\hat{\lambda}^{s}$}\label{large sample property}

Let $\Omega:=\left(a,b,\lambda\right)$ be a given profile of parameter
and in particular, $\Omega_{0}$ be the true parameter. Define
\begin{equation}\label{func V}
\Scale[0.8]{
V_{\Omega}:=p_{a}\left(g\left(Z_{T},T,s\mid a\right),0\right)\mathcal{J}_{Z_T|a}(T)\exp\left(-\int_{0}^{T}\lambda\left(T-s\right)\exp\left(b^{T}g\left(Z_{T},T,s\mid a\right)\right)ds\right)\lambda\left(T\right)
\exp\left(b^{T}Z_{T}\right)
}
\end{equation}
as a random variable with $Z_{T}$ and $T$ being the random variables
following the joint pdf \eqref{joint-pdf} associated with $\Omega_{0}$, denoted as $pdf_{\Omega_0}$. The following technical conditions are needed for the consistency result:

$C1$. The domain $A$ and $B$ are compact. $B$ has open interior with $b_{0}\in B^{\circ}$. The domain of $\lambda$, denoted as $L$, has $\lambda_0\in L$ and is a set of uniformly
bounded right-continuous functions satisfying that $\lambda(0)=1$ for all $\lambda\in L$ (which means $L\subset L^{\infty}\left([0,\infty)\right)$
and under the weak-$\ast$ topology, the closure of $L$ is compact).

$C2$. $E_{\Omega_{0}}\left(\left|\log\left(V_{\Omega}\right)\right|\right)$,
$E_{\Omega_{0}}\left(\left|\nabla_{b_{i}}\log\left(V_{\Omega}\right)\nabla_{b_{j}}\log\left(V_{\Omega}\right)\right|\right)$,
$E_{\Omega_{0}}\left(\left|\nabla_{b_{i}b_{j}}\log\left(V_{\Omega}\right)\right|\right)$
are finite for all $i,j=1,\dots,p$ and all $\Omega\in A\times B\times L$;
and the matrix $I=\left\{ E_{\Omega_{0}}\left(\nabla_{b_{i}}\log\left(V_{\Omega}\right)\nabla_{b_{j}}\log\left(V_{\Omega}\right)\right)\right\} _{1\leq i,j\leq p}$
and $H=\left\{ E_{\Omega_{0}}\left(\nabla_{b_{i}b_{j}}\log\left(V_{\Omega}\right)\right)\right\} _{1\leq i,j\leq p}$
are positive definite.

$C3$. There exists a positive function $d\in L^{1}(pdf_{\Omega_0})-$ such that for all $\Omega\in A\times B\times L$, $\left|\log V_{\Omega}\right|\leq d\left(Z_T,T\right)$ almost surely with respect to the probability measure $pdf_{\Omega_0}$.

$C4$. For all $a\in A$, $q\left(z,t\mid a\right)\in C^{2}\left(\mathbb{R}^{p}\times\mathbb{R}_{+}\right)$
and the map given through $q\left(.\mid.\right):A\rightarrow C^{2}\left(\mathbb{R}^{p}\times\mathbb{R}_{+}\right)$
is continuous with respect to the $C^{2}$ topology. 

$C5$. For every $a\in A$, there is an $p\times p$ matrix $M_a$, such that $\frac{\partial q\left(z,0\mid a\right)}{\partial z}\rightarrow M_a$ as $\left\Vert z\right\Vert\rightarrow \infty$ ($\left\Vert.\right\Vert$ is the Euclidean Norm of a vector). And for different $a$ and $a'$, $M_a-M_{a'}$ has at least one eigenvalue with non-zero real part.

$C6$. $pdf_{\Omega_0}$ has the full support $\mathbb{R}^{p}\times \mathbb{R}_{+}$. The true initial $p(.,0)$ satisfies that $\int_{\mathbb{R}^{p}}\exp(c\cdot z)p(z,0)dz\not=1$ for every $c\in \mathbb{R}^{p}$.
%
%

Condition $C1$-$C3$ are standard for the consistency and asymptotic
normality of maximum likelihood estimators. $C4$ is the regularity
condition that guarantees the trajectories $g\left(.\mid a\right)$
depends on $a$ smoothly. $C5$ is the key to guarantee
the identification of the model \eqref{cox-proportional-hazard} and \eqref{conditional-expectation-of-growth-rate}, although it turns out
that $C5$ can be discarded without any impact on the consistency of $\hat{b}$ and $\hat{\lambda}$, and both of $C5$ and $C6$ can be discarded when the event time and the longitudinal process satisfy a kind of Markovian property and the extra longitudinal observations are available. We will go back to these extensions in the section \ref{extension}. 
\begin{theorem}\label{theorem-identification}
Under Condition $C5$ and $C6$, model \eqref{cox-proportional-hazard} and \eqref{conditional-expectation-of-growth-rate} are identifiable.
And $E_{\Omega_{0}}\left(\log\left(V_{\Omega}\right)\right)$ has
the unique maximal point, $\Omega_{0}$. In addition, if $C4$
holds, $E_{\Omega_{0}}\left(\log\left(V_{\Omega}\right)\right)$
is continuous with respect to the variable $\Omega$.
\end{theorem}

\begin{theorem}\label{theorem-consistency}
(1). Under $C1$-$C6$, the estimator $\hat{a}$, $\hat{b}$ are consistent
and $\hat{b}-b_{0}\rightarrow_{d}N\left(0,I^{-1}\right)$;
 
(2). the estimator
$\hat{\lambda}^{s}$ converges to $\lambda_{0}$ according to the
weak-$*$ topology and $\sqrt{n}\left(\int_{0}^{t}\hat{\lambda}^{s}\left(\tau\right)-\lambda_{0}\left(\tau\right)d\tau\right)$
converges weakly to a Gaussian Process; 

\end{theorem}

\begin{theorem}\label{theorem-ALASSO}
Under $C7$, the estimator $\hat{a}_{AL}$ and $\hat{\lambda}_{AL}^{s}$
has the same properties as $\hat{a}$ and $\hat{\lambda}^{s}$ as
stated in Theorem 2, and the estimator $\hat{b}_{AL}$ has the following oracle property:

(1). denote $\mathcal{A}\subset\left\{ 1,\dots,p\right\} $ as the
set of indices with $b_{0,j}\not=0$ for $j\in\mathcal{A}$ and $\hat{\mathcal{A}}\subset\left\{ 1,\dots,p\right\} $
as the set of indices with $\hat{b}_{0,j}\not=0$ for $j\in\hat{\mathcal{A}}$,
then $\hat{b}_{j}\rightarrow_{p}b_{0,j}$ for all $j\in\mathcal{A}$
and $Prob\left(\hat{\mathcal{A}}=\mathcal{A}\right)\rightarrow1$;

(2). denote $I_{\mathcal{A}}:=\left\{ E_{\Omega_{0}}\left(\nabla_{b_{i}}\log\left(V_{\Omega}\right)\nabla_{b_{j}}\log\left(V_{\Omega}\right)\right)\right\} _{i,j\in\mathcal{A}}$,
$\hat{b}_{L,\mathcal{A}}=\left(\hat{b}_{L,j}\right)_{j\in\mathcal{A}}$
and $b_{0,\mathcal{A}}=\left(b_{0,j}\right)_{j\in\mathcal{A}}$, $\sqrt{n}\left(\hat{b}_{L,\mathcal{A}}-b_{0,\mathcal{A}}\right)\rightarrow_{d}N\left(0,I_{\mathcal{A}}^{-1}\right)$
.
\end{theorem}

\subsection{Extension}\label{extension}

When the longitudinal observations are available at the observation
time before failure occurs at $T$, i.e. the input data has the form $\left\{\left(Z_{i,j},t_{i,j}\right):\,j=1,\dots,m_i;\,i=1,\dots,n\right\}$ with $m_{i}>1$ for $i=1,\dots,n$.
A two-step procedure can be applied to estimate the parameter $\Omega_{0}$, and the resulting estimator turns out to be consistent and have asymptotically normal distribution even without the assumption $C5$ and $C6$. Instead, the following Markovian-style condition are required:
\begin{equation}\label{markovian}
E\left(\epsilon(t-s)\mid Z(t),T\geq t\right)\equiv E\left(\epsilon(t-s)\mid Z(t)\right)
\end{equation}
where $\epsilon(t)$ is the instantaneous variational rate of the longitudinal process as specified in model model \eqref{longitudinal}. Eq. \eqref{markovian} implies  that the conditional mean trajectory that reaches a given realization, $Z(t)$, at the observational time $t$ won't be affected by whether or not the event has already occurred. Formally, the two-step algorithm is stated as following:

Step 1: estimate the parameter $a$ through minimizing the empirical mean of the $L^{2}$
distance between the empirical longitudinal trajectories observed
for each individual $i$ and the theoretical mean trajectories passing
through the point 
$\left(Z_{t_{i,m_{i}}},t_{i,m_i}\right)$:

\begin{equation}\label{mean-L2-distance}
\min_{a\in A}\frac{1}{n}\sum_{i=1}^{n}\frac{1}{m_i}\sum_{j=1}^{m_{i}}\left(g\left(z_{t_{i,m_{i}}},t_{i,m_{i}},t_{i,m_{i}}-t,_{i,j}\mid a\right)-z_{t_{i,j}}\right)^{2}.
\end{equation}
It turns out when 
$n\rightarrow\infty$, the estimator $\hat{a}_{E}$ of solving
the problem \eqref{mean-L2-distance} is consistent. 

Step 2: replace $a$ by $\hat{a}_{E}$
and maximize the likelihood function \eqref{likelihood1} or \eqref{likelihood4} for the parameter $b$ and $\lambda$. 

It turns out that the resulting estimators do have the same properties as stated
in Theorem \ref{theorem-consistency} or \ref{theorem-ALASSO}. The two-step procedure separates the estimation of $a_{0}$ from the
estimation of $b_{0}$ and $\lambda_{0}$. Thanks to this separation, in the second step, the initial pdf and the Jacobian can be completely removed from the likelihood function \eqref{likelihood1} or \eqref{likelihood4} because they only depends the parameter $a$ and the fixed underlying true distribution. In the other words, once if the parameter $a$ is replaced by its first-step estimator $\hat{a}_E$, the component of $\tilde{p}$ becomes constants, and can be deleted from the second-step maximization problem without any impact on the final estimators.  As a consequence, there is no need to calculate the Jacobian $\mathcal{J}_{z_{t_{m_i}}|a}\left(t_{i,m_i}\right)$, which makes the two-step procedure running much faster than the original procedure because the computation of the Jacobian is the most time-consuming part. 

Due to the fact that the conditional density function \begin{equation}
\rho(z,t):=\exp\left(-\int_{0}^{t}\lambda\left(t-s\right)
\exp\left(b^{T}g\left(z,t,s\mid a\right)\right)ds\right)\lambda\left(t\right)
\exp\left(b^{T}z\right)
\end{equation} 
has already had the full support $\mathbb{R}^{p}\times \mathbb{R}_{+}$, and the second-step optimization is equivalent to the optimization of a conditional log-likelihood function formed by the sum of $\log\rho\left(z_{t_{m_i}},t_{m_i}\right)$ for all $m_i$'s, the full support condition in $C6$ can be relaxed.

In addition, from the proof of the theorem \ref{theorem-identification}, it is clear that the main difficulty to achieve the injectivity of the map from the parameter space to the space of all joint pdfs lies in the exclusion of the possibility that different $a$ could lead to the same joint pdf, which is exactly the condition $C5$ and $C6$ designed for. In contrast, when $a_0$ is given, the identification of $b_0$ and $\lambda_0$ becomes trivial and doesn't require any further conditions like $C5$ and $C6$. So when the estimation of $a_0$ can be separated out, $C5$ and $C6$ are redundant. 


Proof of the validity of the two-step procedure is in Appendix \ref{proof-two-step}. A latent assumption behind the proof is that the observational time is uninformative and the total number of observations, $m_i$, is at least $2$ for all subjects. Unlike the joint model discussed in \cite{tsiatis2004joint}, we don't have to assume $m_i$ greater than the dimension of the covariates. This fact makes our two-step procedure more attractive to the scenarios where there are only a few longitudinal observations but a large set of covariates.

\section{Numerical Studies}

\subsection{Simulation Studies}\label{simulation}
In this section, simulation studies are conducted to evaluate the finite-sample performance of the
estimation procedures proposed in section \ref{likelihood function}. Consider the following examples:

\begin{example}\label{example1}
50 samples, each consisting of $n=400,\,800$ subjects, are generated
from simulating the version of model \eqref{potential-growth-process} that has covariate dimension $p=16$, coefficients $b_0=(1,1,-1,0,\dots,0)$ with $3$ non-zero covariate effects. Given $$a=(1,0.5,-1,0.3,1,0.5,-1,0.3,1,0.5,-1,0.3,1,0.5,
-1,0.3),$$ the conditional expectation function \eqref{conditional-expectation-of-growth-rate} is specified as the constant function as below:
\begin{equation}\label{parametrized-conditional-growth-rate}
q\left(z,t\mid a\right)=a.
\end{equation}
The baseline hazard is specified through the function
\begin{equation}\label{lambda0}
\lambda_0(t)=\frac{\exp(10)+\exp^{-t}}{\exp(10)+1}.
\end{equation}
The initial $Z_0\sim N(0,I_{16})$ with $I_{16}$ being the $16$-dimensional identity matrix.
\end{example}


 
The simulation results are presented in the terms of the following criteria:

(1) Figure \ref{fig: fitting} shows the the estimated cumulative hazard $\int_{0}^{t}\hat{\lambda}_{n}^{s}(\tau)d\tau$ versus the true cumulative hazard \ref{lambda0} for Example \ref{example1}. The bias and standard deviation of the estimated non-zero $b_0$ are given in Table \ref{table: non-zero param}. 

(2) We also conduct variable selection for both of the example \ref{example1} by the adaptive LASSO method \eqref{likelihood4}, the result are summarized in Table \ref{table: variable selection} for Example \ref{example1}.
The result are reported by:

i. The average number of the true zero coefficients of $b_0$ that are correctly set to zero, denoted by $C(b_0)$.

ii. The average number of the true non-zero coefficients of $b_0$ that are incorrectly set to zero, which is given by $IC(b_0)$.

iii. The proportion of samples that excluding any non-zero coefficients, denoted by $U-$fit.

iv. The proportion of samples selecting the exact subset models (correct-fit) and the proportion of smaples including all the variables (over-fit), labeled by $C-$fit and $O-$fit respectively.  

From Table \ref{table: non-zero param}, it is clear that for both of the two cases $N=400$ and $N=800$, the fitting to the non-zero coefficients are very good while the fitting accuracy in the case of $N=800$ is even better. As for variable selection, Table \ref{table: variable selection} shows that in most of the samples, the set of zero variables can be exactly identified by our procedure. In particular, as the sample size increases, the identification accuracy is risen up as well. Even in the rare samples where some zero variables are misclassified, the misclassification happened sparsely as $C(b_0)>12$ and that value is close to the true value, $13$.

\subsection{Real Example}\label{real example}
The New York State's Statewide Planning and Research Cooperative System (SPARCS) 2013 is a system initially created to collect information on discharges
from hospitals within New York State. SPARCS currently collects patient
level detail on patient characteristics, diagnoses and treatments,
services, and charges for each hospital inpatient stay and outpatient visit; and each ambulatory surgery and outpatient services visit to
a hospital extension clinic and diagnostic and treatment center licensed to provide ambulatory surgery services. In 2013, the SPARCS contains
nearly 2.5 million inpatient discharges from 218 facilities and 58
counties in New York State. Patient demographics in the SPARCS include
age group at admission, gender,
race, source of payment and zip code. Patient clinical characteristics
include type of admission, diagnosis codes (MDC code, DRG code, CCS
diagnosis code etc.) and treatment procedures undergone (CCS Procedure
Code). 

An important property of the SPARCS data is that there is not any other longitudinal observation available for time-dependent variables, like the cumulative charge, than the observation at the discharge time. Therefore, neither the traditional maximum partial likelihood method nor the estimation procedures designed for the joint models as discussed in \cite{kim2013joint,zeng2007maximum} can be well applied to the SPARCS data. In contrast, the approach proposed in this paper can effectively address the data issue as it is designed for.

In this paper, we consider the discharge time $T$, with the time-dependent covariate, the logarithm of the cumulative charge $Z_1$, and the stationary covariates consisting of the categorical variables, $Z_2,\dots,Z_{25}$, associated with 25 Major Diagnosis Code (MDC) and the degree ($1\sim 4$) of the Severity of Illness, $Z_{26}$. Our analysis is conducted on a subsample of the entire SPARCS 2013 database with sample size $400$. The summary statistics of our subsample are presented in Table \ref{table: 3.1}.

The penalized maximum likelihood estimators $\hat{b}_{AL}$ are reported in Table \ref{table: real}. The non-parametric estimator $\int_0^t\hat{\lambda}^{s}_{n}(s)ds$ for the cumulative baseline hazard are plotted in Figure \ref{fig: real data study}.

In Table \ref{table: real}, the significant negative coefficients for log-charge indicates the strong positive correlation between the total charge and los. In addition, it seems that there does not exist robust connection between the los and the severity/mortality of illness. 

By Figure \ref{fig: real data study}, the day 5 seems to be relatively special because the variation of the slope of the cumulative hazard turns from increasing to decreasing around this time, which implies that for patients who have already stayed in hospital for 5 days, they are more probable to have a longer stay.

\section{Remarks and Conclusion}\label{conclusion}
In this paper, we proposed a maximum full information likelihood procedure to estimate the Cox model with temporal covariates. The most significant advantage of our procedure is that it can generate well-performed estimation without requiring the extra longitudinal observations before the time to event. There are also  three potential extensions to the current work.  

\subsection{Censoring}
Although censoring is not discussed in the current framework, it can be added in the standard way such that censoring is (1) independent from the occurrence of the interested event, or (2) conditionally independent from the event given the covariates at the observational time. In both of the two cases, the consistency and asymptotic normality of the resulting estimators still hold and their proof is straightforward from the proof of Theorem \ref{theorem-consistency} and \ref{theorem-ALASSO}.

\subsection{Forecast Long Term Survival Rate}
In addition to the hazard function \eqref{cox-proportional-hazard}, the estimators proposed in section \ref{likelihood function} indicates a consistent estimator to the long term survival rate (LTSR):
\begin{equation}\label{LTSR}
S\left(z,t,t'\right):=Prob\left(T\in \left[t,t+t'\right)\mid Z_t=z\right).
\end{equation}
Using the notation $g^{-1}$ as in Eq. \eqref{g-inverse}, the estimator to \eqref{LTSR} can be given as below:
\begin{equation}\label{hat-LTSR}
\hat{S}\left(z,t,t'\right):=\exp\left(
-\int_{t}^{t+t'}\exp\left(\hat{b}^{T}g^{-1}\left(
z,t,\tau\mid \hat{a}\right)\right)\hat{\lambda}_n^s
\left(\tau\right)d\tau\right),
\end{equation}
where $\hat{a}$, $\hat{b}$ and $\hat{\lambda}_n^s$ are the estimators derived in section \ref{likelihood function}, which can be replaced by their penalized version as well. The consistency and asymptotic normality of the estimator \eqref{hat-LTSR} is just a direct result of the theorem \ref{theorem-consistency} and/or \ref{theorem-ALASSO}. It is worthwhile to mention that \eqref{hat-LTSR} is not possible to be constructed from the maximum partial likelihood estimators of the Cox model when temporal covariates are included. Because it is clear from \eqref{hat-LTSR} that $\hat{S}$ relies on the information of the temporal covariates $Z$ within the forecast interval $\left[t,t+t'\right)$, which is not available from the maximum partial likelihood estimators.

\subsection{Semi-martingale Longitudinal Processes}
Although in the current discussion, the longitudinal process is assumed to have bounded variation and absolutely continuous with respect to the Lebesgue measure on $\mathbb{R}_+$, the same framework should be extensible to more general cases where the longitudinal process may not have bounded variation (for example, given by a semi-martingale process). In a series of related works, the authors construct an explicit expression of the joint pdf of the event time and a semi-martingale longitudinal process, which enables us to construct the full information likelihood function. But the challenges to extend the current work to the situation with the semi-martingale longitudinal measurements are the identification of the resulting model and the challenge in computation. For the identification issue, it is clear from the proof \ref{proof-identification} that in the current framework, the identification relies on detailed analysis of the solution trajectories of ODE system induced by the function \eqref{conditional-expectation-of-growth-rate}. In the case of semi-martingale longitudinal measurements, the ODE system will be replaced with a more complicated partial differential equation (PDE) system. Although it seems that there is no barrier to make the same trick in proof \ref{proof-identification} invalid, the details to transplant the proof \ref{proof-identification} to the semi-martingale case is open to future studies. In the aspect of computation, we have to apply numerical method to a PDE system in place of an ODE system, while,as known, the numerical method to solve PDE system is much more time consuming. A potential solution to the computation issue is to utilize the relation between PDE systems and the semi-martingale processes, through which simulating the underlying process could yield exactly the same solution to the PDE problem. The details of implementing that idea are left as another open problem for further study.

\begin{appendices}
\section{}

\subsection{Proof for Theorem \ref{theorem-identification}}\label{proof-identification}

The identifiability of the model \eqref{cox-proportional-hazard} and \eqref{conditional-expectation-of-growth-rate} is equivalent to that
as long as $\Omega\not=\Omega_0$, 
\begin{equation}\label{unique-maximum}
\Scale[0.9]{
\begin{aligned}
&p_{a}\left(g\left(z,t,t\mid a\right),0\right)\mathcal{J}_{z|a}(t)\exp(-\int_{0}^{t}\exp(b^Tg\left(z,t,s\mid a\right))\lambda(t-s)ds)\lambda(t)\exp\left(b^{T}z\right)\\
&\not=
p_{a_0}\left(g\left(z,t,t\mid a_0\right),0\right)\mathcal{J}_{z|a_0}(t)\exp(-\int_{0}^{t}\exp(b_0^{T}g\left(z,t,s\mid a_0\right))\lambda_0(t-s)ds)\lambda_0(t)\exp\left(b_0^{T}z\right)
\end{aligned}}
\end{equation}
within a positive measure set $M\in\mathbb{R}^{p}\times\mathbb{R}_{+}$
(with respect to the standard Lebesgue Measure).

(1) Suppose \eqref{unique-maximum} does not hold for some $\Omega$ with $b\not=b_0$. The right-continuity condition in $C1$ requires that
\begin{equation}\label{initial identity}
p_a(z,0)\exp(b^Tz)\equiv p(z,0)\exp(b_0^Tz)
\end{equation}
where $p(.,0)$ is the true initial pdf. The assumption that the true joint pdf \eqref{joint-pdf} has the full support implies that $p(.,0)$ has full support as well. Therefore, Eq. \eqref{initial identity} leads to 
\begin{equation}
p_a(z,0)\equiv p(z,0)\exp\left(\left(b_0-b\right)^Tz\right),
\end{equation} 
both $p_a(.0)$ and $p(.,0)$ are probability density function, which yield that 
\begin{equation}
\int_{\mathbb{R}^{p}}p(z,0)\exp(c\cdot z)dz=1
\end{equation}
for some $c\not=0$ that contradicts to the requirement in $C6$. Consequently, every $\Omega$ that could potentially break down the condition \eqref{unique-maximum} must have $b=b_0$. 

(2) On the other hand, if $b=b_0$, we have for all $t\geq 0$:
\begin{equation}\label{condition-for-lambda}
S_{(a,b_0,\lambda)}(t)\cdot \lambda(t)=S_{(a_0,b_0,\lambda_0)}(t)\cdot \lambda_0(t)
\end{equation}
where $S_{(a,b,\lambda)}$ is the survival function of the failure time that follows the joint pdf associated with $\Omega=(a,b,\lambda)$, by definition it has the following form: 
\begin{equation}
\Scale[0.9]{
S_{(a,b,\lambda)}(t):\int_{\mathbb{R}^p}p_{a}\left(g\left(z,t,t\mid a\right),0\right)\mathcal{J}_{z|a}(t)\exp(-\int_{0}^{t}\exp(b_0^Tg\left(z,t,s\mid a\right))\lambda(t-s)ds)dz\lambda(t)}.
\end{equation}
Because the survival function is uniquely determined by the pdf of the event time which is furthermore uniquely determined by the joint pdf. Under the assumption that $\Omega$ and $\Omega_0$ corresponds to exactly the same joint pdf, the equation \eqref{condition-for-lambda} enforces that $\lambda=\lambda_0$ for all $\Omega$ that breaks the condition \eqref{unique-maximum}.

(3) Suppose there exists $\Omega=(a,b_0,\lambda_0)$ with $a\not=a_0$ for which the condition \eqref{unique-maximum} doesn't hold. Then, the following identity holds:
\begin{equation}\label{weak-identity}
\Scale[0.9]{
\begin{aligned}
&p_{a}\left(z,0\right)\mathcal{J}_{g^{-1}\left(z,0,t\mid a\right)|a}(t)\exp(-\int_{0}^{t}\exp(b^Tg^{-1}\left(z,0,s\mid a\right))\lambda(s)ds)\\
&=
p_{a_0}\left(z_0,0\right)\mathcal{J}_{g^{-1}\left(z_0,0,t\mid a_0\right)|a_0}(t)\exp(-\int_{0}^{t}\exp(b_0^{T}g^{-1}\left(z_0,0,s\mid a_0\right))\lambda_0(s)ds).
\end{aligned}}
\end{equation}
for all pairs $(z,z_0)$ such that $z_0=g\left(g^{-1}\left(z,0,t\mid a\right),t,t\mid a_0\right)$ where $g^{-1}\left(z,s,t\mid a\right)$ is
the inverse trajectories of $g$ and defined through the relation
\begin{equation}\label{g-inverse}
g\left(g^{-1}\left(z,s,t\mid a\right),s+t,t\mid a\right)=z.
\end{equation} Factor out Eq. \eqref{weak-identity} by $\mathcal{J}_{g^{-1}\left(z_0,0,t\mid a_0\right)|a_0}(t)\exp(-\int_{0}^{t}\exp(b_0^{T}g^{-1}\left(z_0,0,s\mid a_0\right))\lambda_0(s)ds)$ and take the limit as $t\rightarrow 0$ yielding the following identity:
\begin{equation}\label{invariant-measure}
p_a(\mathcal{T}_{r}(z_0),0)\cdots\mathcal{J}_{\mathcal{T}_r}(z_0)=p(z,0)
\end{equation} 
where for every $r\in \mathbb{R}$, the map $\mathcal{T}_{r}:\mathbb{R}^{p}\rightarrow \mathbb{R}^{p}$ is the diffeomorphism obtained from solving the ODE system:
\begin{equation}\label{ode2}
z'=q\left(z,0\mid a\right)-q\left(z,0\mid a_0\right),
\end{equation}
$\mathcal{T}_r(z_0)$ is just the point reached at the time $r$ by the trajectory starting at $z_0$ that solves Eq. \eqref{ode2}. $\mathcal{J}_{\mathcal{T}_r}$ is the Jacobian associated with $\mathcal{T}_r$. By the language of ergodic theory, Eq. \eqref{invariant-measure} implies that the probability measure $p(.,)$ is invariant under the $\mathbb{R}$-action on the space $\mathbb{R}^{p}$ induced by the solutions $\mathcal{T}$. However, under the condition $C5$, the action $\mathcal{T}$ associated with the pair of $a$ and $a_0$ does not allow any invariant probability measure fully supported on $\mathbb{R}^{p}$ unless $a=a_0$. This contradiction guarantees the condition \eqref{unique-maximum}.

The uniqueness of the maximal point $\Omega_{0}$ of function $E_{\Omega_{0}}\left(\log\left(V_{\Omega}\right)\right)$
is simply the consequence of the standard proof the consistency of the full information maximum likelihood estimator, and can be found in every advanced textbook of econometrics, like \cite{amemiya1985advanced}.

%
%
%
The continuity of $E_{\Omega_{0}}\left(\log\left(V_{\Omega}\right)\right)$
with respect to $\Omega$ and its differentiability with respect to
the component $b$ comes from $C4$ by the the dominant
convergent theorem. This completes the proof for Theorem \ref{theorem-identification}.

\subsection{Proof for Theorem \ref{theorem-consistency}}\label{proof-consistency}

The relation \eqref{constraint-for-lambda} defines a map, denoted as $\hat{\iota}_n$, that assigns every $(a,b)\in A\times B$ a step-wise function with the step heights specified by \eqref{constraint-for-lambda}. We can define the asymptotic version of $\hat{\iota}_n$ as below:
\begin{equation}
\iota\left(a,b\right)(t):=
\frac{\int_{\mathbb{R}^{p}}pdf_{\Omega_{0}}
\left(z,t\right)dz}{\int_{t}^{\infty}\int_{\mathbb{R}^{p}}pdf_{\Omega_{0}}
\left(z,\tau\right)\cdot\exp\left(b^T g\left(z,\tau,\tau-t\mid a\right)\right)dzd\tau},
\end{equation}
where $pdf_{\Omega_0}$ is the joint pdf associated with the true parameter $\Omega_0$. It turns out that $\iota$ is continuous with respect to the weak$-\ast$ topology on the $L^{\infty}$ space and has the compact domain $A\times B$. In addition, for every pair $(a,b)$, $\hat{\iota}_n(a,b)\rightarrow \iota(a,b)$ in the weak$-\ast$ topology. 

Then, the consistency of $\hat{a}$, $\hat{b}$ and $\hat{\lambda}^{s}$
follows immediately from $C2$ and the facts: (1) the function $l_n\left(a,b,\hat{\iota}_n\left(a,b\right)\right)\rightarrow_p E_{\Omega_{0}}\left(\log\left(V_{\left(a,b,\iota\left(a,b\right)\right)}\right)\right)$ (by the strong law of large number); (2) the function $E_{\Omega_{0}}\left(\log\left(V_{\left(a,b,\iota\left(a,b\right)\right)}\right)\right)$ is continuous and
has a unique maximal point, $\Omega_{0}$, within a compact domain (by the theorem \ref{theorem-identification}).

The asymptotic normality of $\hat{b}$ can be verified by the standard
argument for the asymptotic normality of a maximum likelihood estimator.

To verify the asymptotic normality of $\hat{\lambda}^{s}$, firstly
notice that let $\left\{ Z_{t}:t\in[0,\infty)\right\} $ be a process
satisfying model \eqref{conditional-expectation-of-growth-rate} associated with the true parameter $a_{0}$ and
$\left(Z_{T},T\right)$ be a random vector following the distribution
associated with $\Omega_{0}$, denote $N\left(t\right):=I\left(T\leq t\right)$
being the counting process determined by $T$ and $\tilde{N}\left(t\right):=I\left(T> t\right)$. The processes $N\left(t\right)$
and $\tilde{N}(t)$ determines a martingale process as below:

\begin{equation}\label{estimator-process}
M\left(t\right):=\int_{0}^{t}\frac{dN\left(s\right)}{E\left(\exp\left(b_{0}^{T}Z_{s}\right)\cdot \tilde{N}\left(s\right)\right)}-\frac{
\tilde{N}\left(t\right)}{E\left(\tilde{N}(t)\right)}\int_{0}^{t}\lambda_{0}\left(\tau\right)d\tau
\end{equation}
it turns out that $E\left(M\left(t\right)\right)\equiv0$,
\[
Var\left(M\left(t\right)\right)=\int_{0}^{t}
\frac{\lambda_{0}\left(s\right)}{E\left(\exp\left(b_{0}^{T}Z_{s}\right)\cdot \tilde{N}\left(s\right)\right)}ds+\frac{\left(\int_{0}^{t}
\lambda_{0}\left(s\right)ds\right)^{2}}{E\left(\tilde{N}(t)\right)}\]
for $t>s$. On the other hand, by the relation in Eq. \eqref{constraint-for-lambda}, we have:

\begin{equation}
\Scale[0.8]{
\begin{aligned}
\sqrt{n}\left(\int_{0}^{t}\hat{\lambda}^{s}\left(\tau\right)-\lambda_{0}\left(\tau\right)d\tau\right)  &=  \frac{1}{\sqrt{n}}\left(
\begin{aligned}
&\sum_{i=1}^{j_{t}}\frac{n\cdot\left(t_{i}-t_{i-1}\right)}{\sum_{j=i+1}^{n}\int_{0}^{t_{i}-t_{i-1}}\exp\left(b^{T}g\left(z_{t_{j}},
t_{j},t_{j}-t_{i}+\tau|a\right)\right)d\tau}\\
&+\frac{n\cdot\left(t-t_{j_{t}}\right)}{\sum_{j=j_t+1}^{n}\int_{0}^{t_{j_{t}}-t_{j_{t}-1}}\exp\left(b^{T}g\left(z_{t_{j}}
,t_{j},t_{j}-t_{j_t}+\tau|a\right)\right)d\tau}-\frac{n-j_t}{\frac{n-j_t}{n}}\int_{0}^{t}\lambda_{0}\left(\tau\right)d\tau
\end{aligned}
\right)\\
 & =  \frac{1}{\sqrt{n}}\cdot\sum_{i=1}^{n}\left(\frac{I\left(t_{i}<t\right)}{\frac{\sum_{j=i+1}^{n}\exp\left(b^{T}g\left(z_{t_{j}},
t_{j},t_{j}-t_{i}|a\right)\right)}{n}}-\frac{I\left(t_{i}\geq t\right)}{\frac{n-j_t}{n}}\int_{0}^{t}\lambda_{0}\left(\tau\right)d\tau\right)+
 O\left(h_n\right)\\
 & =  \frac{\sum_{i}^{n}M_{i}\left(t\right)}{\sqrt{n}}+O\left( h_n\right)
\end{aligned}}
\end{equation}
where $j_{t}=\max\left\{ i\in\left\{ 1,\dots,n\right\} :t_{i}<t\right\} $ and 
$h_n=\max\left(h_n^1,h_n^2,h_n^3\right)$, with $h_n^1$, $h_n^2$ and $h_n^3$ given as below:
\begin{equation*}
\begin{aligned}
h_n^1 &:=\sup\left\{ t_{i}-t_{i-1}:i=1,\dots,n\right\}\\
h_n^2 &:=\sup\left\{\left|\frac{\left(n-i\right)}{n}-E\left(\tilde{N}\left(t\right)\right)\right|:\, i\leq j_{t}\right\}\\
h_n^3 &:=\Scale[0.9]{\sup\left\{\left|\frac{\sum_{j=i+1}^{n}\exp\left(b^{T}g\left(z_{t_{j}},
t_{j},t_{j}-t_{i}|a\right)\right)}{n}-E\left(\exp\left(b_{0}^{T}Z_{s}\right)\cdot \tilde{N}\left(s\right)\right)\right|:i\leq j_t\right\}}
\end{aligned}
\end{equation*}
By the assumption that $\lambda_{0}$ is strictly positive, the fact that $\sup\left\{ t_{i}-t_{i-1}:i=1,\dots,n\right\} \rightarrow_{p}0$, and that for every fixed $t<\infty$, $\left|\frac{\left(n-i\right)}{n}-E\left(1-N\left(t\right)\right)\right|\rightarrow_p 0$, $\Scale[0.75]{\left|\frac{\sum_{j=i+1}^{n}\exp\left(b^{T}g\left(z_{t_{j}},
t_{j},t_{j}-t_{i}|a\right)\right)}{n}-E\left(\exp\left(b_{0}^{T}Z_{s}\right)\cdot \tilde{N}\left(s\right)\right)\right|\rightarrow_p 0}$ uniformly by uniform law of large number. Therefore, by central limit
theorem, we have:

\begin{equation}\label{limit-eq1}
\lim_{n\rightarrow\infty}\sqrt{n}\left(\int_{0}^{t}\hat{\lambda}^{s}\left(\tau\right)-\lambda_{0}\left(\tau\right)d\tau\right)=\lim_{n\rightarrow\infty}\frac{\sum_{i}^{n}\left(M_{i}\left(t\right)-\int_{0}^{t}\lambda_{0}\left(\tau\right)d\tau\right)}{\sqrt{n}}=N\left(0,Var\left(M\left(t\right)\right)\right)
\end{equation}
Applying a vector version of the central limit theorem as well as
Eq. \eqref{limit-eq1}, we can prove the weak convergence of $\sqrt{n}\left(\int_{0}^{t}\hat{\lambda}^{s}\left(\tau\right)-\lambda_{0}\left(\tau\right)d\tau\right)$
to the Gaussian Process $B\left(Var\left(M\left(t\right)\right)\right)$.

Finally, the consistency and the weak convergence of the estimator
$_{t}\hat{S}_{t'}$ is direct from the consistency and asymptotic
normality of $\hat{\lambda}^{s}$.

\subsection{Proof for Theorem \ref{theorem-ALASSO}}\label{proof-alasso}

Construct a function $\Psi_{n}\left(a,u,\lambda^{s}\right)$ with
$u\in\mathbb{R}^{p}$ as below:

\begin{equation}\label{Psi-n}
\Psi_{n}\left(a,u,\lambda^{s}\right):=l_{n}\left(a,\hat{b}+u,\lambda^{s}\right)-\Lambda_{n}\sum_{i=1}^{p}\frac{\left|\hat{b}_{j}+\frac{u_{j}}{\sqrt{n}}\right|}{\left|\hat{b}_{j}\right|^{2}}
\end{equation}

Define $W_{n}\left(a,u,\lambda^{s}\right):=\Psi_{n}\left(a,u,\lambda^{s}\right)-\Psi_{n}\left(a,0,\lambda^{s}\right)$,
then $W_{n}$ has the following form:
\begin{equation}\label{W-n}
W_{n}\left(a,u,\lambda^{s}\right) = l_{n}\left(a,\hat{b}+u,\lambda^{s}\right)-l_{n}\left(a,\hat{b},\lambda^{s}\right)-\frac{\Lambda_{n}}{\sqrt{n}}\sum_{i=1}^{p}\frac{\sqrt{n}\left(\left|\hat{b}_{j}+\frac{u_{j}}{\sqrt{n}}\right|-\left|\hat{b}{}_{j}\right|\right)}{\left|\hat{b}{}_{j}\right|^{2}}.
\end{equation}

If $b_{0,j}\not=0$, then $\hat{b}_{j}\rightarrow_{p}b_{0,j}$ and
$\sqrt{n}\left(\left|\hat{b}_{j}+\frac{u_{j}}{\sqrt{n}}\right|-\left|\hat{b}_{j}\right|\right)\rightarrow u_{j}\cdot\frac{b_{0,j}}{\left|b_{0,j}\right|}$.
Therefore we have $\frac{\Lambda_{n}}{\sqrt{n}}\frac{\sqrt{n}\left(\left|\hat{b}_{j}+\frac{u_{j}}{\sqrt{n}}\right|-\left|\hat{b}_{j}\right|\right)}{\left|\hat{b}_{j}\right|^{2}}\rightarrow_{p}0$ 
by the assumption $\frac{\Lambda_{n}}{\sqrt{n}}\rightarrow0$. If
$b_{0,j}=0$, $\sqrt{n}\left(\left|\hat{b}_{j}+\frac{u_{j}}{\sqrt{n}}\right|-\left|\hat{b}_{j}\right|\right)=\left|u_{j}\right|$
and $\frac{\Lambda_{n}}{\sqrt{n}\left|\hat{b}_{j}\right|}=\sqrt{n}\Lambda_{n}\frac{1}{\left|\sqrt{n}\hat{b}_{j}\right|^{2}}$,
as $\sqrt{n}\hat{b}_{j}=O\left(1\right)$ as $n\rightarrow\infty$
almost surely, we have $\frac{\Lambda_{n}}{\sqrt{n}\left|\hat{b}_{j}\right|}\rightarrow_{p}\infty$
by the assumption $\sqrt{n}\Lambda_{n}\rightarrow\infty$. Consequently,

\begin{equation}\label{W}
\Scale[0.8]{
W_{n}\left(a,u,\lambda^{s}\right)\rightarrow_{p}W\left(a,u,\lambda^{s}\right)=\begin{cases}
E_{\Omega_{0}}\left(\log\left(V_{\left(a,b_{0}+u,\lambda^{s}\right)}\right)\right)-E_{\Omega_{0}}\left(\log\left(V_{\Omega_{0}}\right)\right) & \textrm{if }u_{j}=0\textrm{ for all }j\in\mathcal{A}\\
-\infty & \textrm{else}
\end{cases}}
\end{equation}
which implies all the estimators, $\hat{a}_{AL}$, $\hat{b}_{AL}$ and
$\hat{\lambda}_{AL}^{s}$, converges to the unique maximal points of
the function $E_{\Omega_{0}}\left(\log\left(V_{\left(a,b_{0}+u,\lambda^{s}\right)}\right)\right)-E_{\Omega_{0}}\left(\log\left(V_{\Omega_{0}}\right)\right)$,
which is the true value $a_{0}$, $b_{0}$ and $\lambda_{0}$. The
asymptotic normality of $\sqrt{n}\left(\hat{\lambda}_{AL}^{s}-\lambda_{0}\right)$
is established in the same way as in the proof of Theorem \ref{theorem-consistency}. The oracle
property of $\hat{b}_{AL}$ can be verified as below:

First, it is obvious that for each $j\in\mathcal{A}$, $\lim_{n\rightarrow\infty}Prob\left(j\in\hat{\mathcal{A}}\right)=1$.
And for $j\not\in\mathcal{A}$, if $j\in\hat{\mathcal{A}}$, by the
first order condition of the maximization problem \eqref{likelihood4}, we have the
identity that:

\begin{equation}\label{identity-of-l-n}
\left|\frac{\partial l_{n}}{\partial_{u_{j}}}\right|=\frac{\Lambda_{n}}{\sqrt{n}\left|\hat{b}_{j}\right|^{2}}=\frac{\sqrt{n}\Lambda_{n}}{\left|\sqrt{n}\hat{b}_{j}\right|^{2}}
\end{equation}
where the right-hand side $\rightarrow_{p}\infty$ while the left-hand
side $\rightarrow_{p}\frac{\partial E_{\Omega_{0}}\left(\log\left(V_{\Omega_{0}}\right)\right)}{\partial_{b_{j}}}=0$
, while implies that $Prob\left(j\not\in\mathcal{A},j\in\hat{\mathcal{A}}\right)\leq Prob\left(\left|\frac{\partial l_{n}}{\partial u_{j}}\right|=\frac{\Lambda_{n}}{\sqrt{n}\left|\hat{b}_{j}\right|^{2}}\right)\rightarrow0$.
It verifies that $\lim_{n\rightarrow\infty}Prob\left(\hat{\mathcal{A}}=\mathcal{A}\right)=1$.

Second, the asymptotic normality of $\hat{b}_{AL,\mathcal{A}}$. Consider 

\begin{equation}\label{derivative-identity-for-Psi-n}
\nabla_{u_{\mathcal{A}}}\Psi_{n}\left(a,0,\lambda^{s}\right) =  -\nabla_{u_{\mathcal{A}}}\nabla_{u_{\mathcal{A}}}\Psi_{n}\left(a,u',\lambda^{s}\right)u_{\mathcal{A}}
\end{equation}
where the left-hand side equals to $\nabla_{u_{\mathcal{A}}}l_{n}\left(a,\hat{b},\lambda^{s}\right)-\left(\frac{\Lambda_{n}}{\sqrt{n}\left|\hat{b}_{j}\right|^2}\right)_{j\in \mathcal{A}}$
with $\sqrt{n}\left(\frac{\Lambda_{n}}{\sqrt{n}\left|\hat{b}_{j}\right|^2}\right)_{j\in \mathcal{A}}\rightarrow0$
and the right-hand side equals to $-\nabla_{u_{\mathcal{A}}}\nabla_{u_{\mathcal{A}}}l_{n}\left(a,u',\lambda^{s}\right)u_{\mathcal{A}}$
with $u'$ being some intermediate point between $0$ and $\hat{b}_{AL}-\hat{b}$.
By the condition $C2$ and the consistency property of estimator $\hat{b}$,
we have 

\begin{equation}\label{normality-conclusion-ALASSO}
\sqrt{n}\left(\hat{b}_{AL,\mathcal{A}}-\hat{b}\right)=\sqrt{n}u_{\mathcal{A}}\sim
\nabla_{u_{\mathcal{A}}}\nabla_{u_{\mathcal{A}}}
l_{n}\left(a,\hat{b}+u',\lambda^{s}\right)^{-1}\cdot\sqrt{n}
\nabla_{u_{\mathcal{A}}}l_{n}\left(a,\hat{b},\lambda^{s}\right)\rightarrow_{d}N\left(0,I_{\mathcal{A}}^{-1}\right).
\end{equation}

\subsection{Proof for the Validity of the Two-Step Procedure}\label{proof-two-step}

It to show the estimator $\hat{a}_E$ from the first step is consistent, which is equivalent to show that the following function has a unique minimum:
\begin{equation}\label{mean_dis}
m(a):=E\left(E\left(\int_{0}^{t}\left(Z(s)-g\left(z,t,t-s\mid a\right)\right)^{2}ds\mid T=t,Z(t)=z\right)\right).
\end{equation}
In fact, the unique minimal point of the function \eqref{mean_dis} must be $a_0$ as long as $g\left(z,t,t-s\mid a\right)$ equals to the conditional mean of $E\left(Z(s)\mid Z(t)=z,T=t\right)$ for all $z$ and $s\leq t$, which is implied by the condition \eqref{markovian}.

\section{Tables \& Figures}
\subsection{Tables}
\vspace{0.5cm}
\begin{minipage}{\linewidth}
\begin{center} 
\captionof{table}{Descriptive statistics of SPARCS Sample}\label{table: 3.1}
\centering
\resizebox{8cm}{5cm}{%
\begin{tabular}{lllll}
\hline
\hline
Characteristics & Group & Charge(SD) & N(\%) & LOS(SD)\tabularnewline
\hline
All Patients &  & 9.86(1.05) & 400(100) & 5.25(7.44)\tabularnewline
MDC & 1 & 10.2(0.83) & 19(4.76) & 5.26(5.28)\tabularnewline
 & 3 & 9.86(0.62) & 7(1.5) & 2.67(2.34)\tabularnewline
 & 4 & 10.05(1.08) & 27(6.77) & 6.07(6.66)\tabularnewline
 & 5 & 10.31(1.03) & 55(13.78) & 4.58(4.88)\tabularnewline
 & 6 & 10.01(0.83) & 35(8.77) & 4.71(4.5)\tabularnewline
 & 7 & 10.33(0.95) & 10(2.51) & 4.9(2.88)\tabularnewline
 & 8 & 10.51(0.69) & 25(6.27) & 4.84(3.29)\tabularnewline
 & 9 & 9.93(0.84) & 10(2.51) & 6.7(6.27)\tabularnewline
 & 10 & 9.92(0.41) & 12(3.01) & 3.0(1.81)\tabularnewline
 & 11 & 10.25(0.98) & 14(3.51) & 6.0(4.37)\tabularnewline
 & 12 & 10.16 & 1(0.25) & 1\tabularnewline
 & 13 & 10.01(0.89) & 7(1.75) & 1.86(1.21)\tabularnewline
 & 14 & 9.39(0.55) & 44(11.03) & 2.39(0.78)\tabularnewline
 & 15 & 8.66(0.86) & 50(12.53) & 3.26(5.29)\tabularnewline
 & 16 & 10.19(1.12) & 5(1.25) & 6.0(4.8)\tabularnewline
 & 17 & 10.03(0.92) & 5(1.25) & 4.8(2.95)\tabularnewline
 & 18 & 10.07(1.38) & 17(4.26) & 10.59(22.32)\tabularnewline
 & 19 & 9.93(1.25) & 24(6.02) & 12.42(12.8)\tabularnewline
 & 20 & 9.55(0.94) & 15(3.76) & 5.47(6.88)\tabularnewline
 & 21 & 10.61(1.13) & 5(1.25) & 8.6(11.17)\tabularnewline
 & 22 & 11.8 & 1(0.25) & 15\tabularnewline
 & 23 & 9.9(0.91) & 9(2.26) & 8.44(5.13)\tabularnewline
 & 24 & 10.0(1.01) & 3(0.75) & 3.33(1.53)\tabularnewline
Severity & 0 & 9.51(0.93) & 250(62.66) & 3.88(5.75)\tabularnewline
 & 1 & 10.23(0.9) & 80(20.05) & 5.46(4.23)\tabularnewline
 & 2 & 10.54(0.97) & 43(10.78) & 8.07(5.61)\tabularnewline
 & 3 & 10.87(1.19) & 26(6.52) & 13.12(18.65)\tabularnewline
Mortality & 1 & 9.38(0.87) & 159(39.85) & 3.14(5.38)\tabularnewline
 & 2 & 9.93(1.0) & 137(34.34) & 5.05(5.47)\tabularnewline
 & 3 & 10.38(0.95) & 79(19.8) & 7.22(5.39)\tabularnewline
 & 4 & 10.95(1.12) & 24(6.02) & 13.88(19.09)\tabularnewline
 \hline
 \hline
\end{tabular}%
}
\end{center}
\end{minipage}\\

\vspace{2cm}
\begin{minipage}{\linewidth}
\begin{center} 
\centering
\captionof{table}{Estimation Bias and Std.}\label{table: non-zero param}
\resizebox{12cm}{1.2cm}{%
\begin{tabular}{lcccc}
\hline
\hline
 & bias($N=400$) & std($N=400$) & bias($N=800$) & std($N=800$)\tabularnewline
 \hline
$b_1$  & -0.002 & 0.052 & -0.004 & 0.049\tabularnewline
$b_2$ & -0.011 & 0.041 & -0.007 & 0.048\tabularnewline
$b_3$ & -0.003 & 0.046 & 0.001 & 0.052 \tabularnewline
\hline
\hline
\end{tabular}%
}
\end{center}
\end{minipage}\\

\vspace{1cm}

\begin{minipage}{\linewidth}
\begin{center} 
\centering
\captionof{table}{Estimation Bias and Std.}\label{table: variable selection}
\resizebox{11cm}{1cm}{%
\begin{tabular}{cccccc}
\hline
\hline
Sample Size & $C(b_0)$ & $IC(b_0)$ & $U\_fit$ & $C\_fit$ & $O\_fit$\tabularnewline
\hline
$N=400$ & 12.59 & 0 & 0 & 0.69 & 0\tabularnewline
$N=800$ & 12.53 & 0 & 0 & 0.76 & 0\tabularnewline
\hline
\hline
\end{tabular}%
}
\end{center}
\end{minipage}\\

\vspace{2cm}

\begin{minipage}{\linewidth}
\begin{center} 
\centering
\captionof{table}{Estimation Coefficients for Real Example}\label{table: real}
\resizebox{4cm}{5cm}{%
\begin{tabular}{lc}
\hline
\hline
Variables & Estimated Values\tabularnewline
\hline
Log Charge & -0.79\tabularnewline
MDC1 & 0.16\tabularnewline
MDC2 & -0.14\tabularnewline
MDC3 & 0.2\tabularnewline
MDC4 & 0\tabularnewline
MDC5 & 0\tabularnewline
MDC6 & 0.1\tabularnewline
MDC7 & 0\tabularnewline
MDC8 & 0\tabularnewline
MDC9 & 0\tabularnewline
MDC10 & 0.15\tabularnewline
MDC11 & 0\tabularnewline
MDC12 & -0.09\tabularnewline
MDC13 & -0.28\tabularnewline
MDC14 & 0.22\tabularnewline
MDC15 & 0\tabularnewline
MDC16 & 0\tabularnewline
MDC17 & 0\tabularnewline
MDC18 & 0.26\tabularnewline
MDC19 & 0\tabularnewline
MDC20 & 0\tabularnewline
MDC21 & 0\tabularnewline
MDC22 & 0\tabularnewline
MDC23 & 0\tabularnewline
MDC24 & 0\tabularnewline
Severity & 0\tabularnewline
Mortality & 0\tabularnewline
\hline
\hline
\end{tabular}%
}
\end{center}
\end{minipage}\\
\subsection{Figures}
\vspace{0.5cm}
\begin{minipage}\linewidth
\begin{center}
\includegraphics[width=15cm,height=7cm]{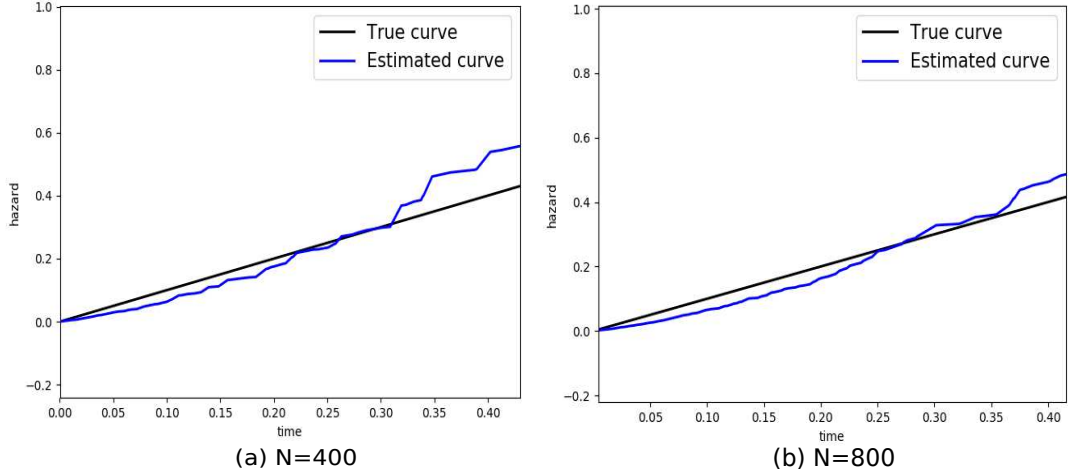}
\captionof{figure}{Estimated $\int_{0}^{t}\hat{\lambda}^s_n(\tau)d\tau$ v.s. True $\int_{0}^{t}\lambda(\tau)d\tau$}\label{fig: fitting}
%
\end{center}
\end{minipage}\\
\vspace{1cm}

\begin{minipage}\linewidth
\begin{center}
\includegraphics[width=7cm,height=4.5cm]{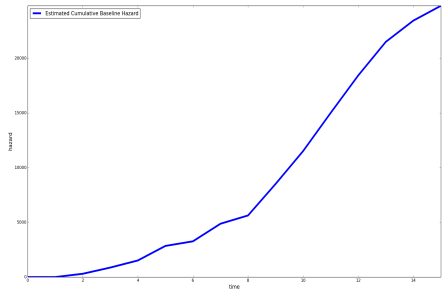}
\captionof{figure}{Estimated Cumulative Baseline Hazard for Real Sample}\label{fig: real data study}
%
\end{center}
\end{minipage}\\
\end{appendices}

\begin{thebibliography}{20}
\providecommand{\natexlab}[1]{#1}
\providecommand{\url}[1]{\texttt{#1}}
\expandafter\ifx\csname urlstyle\endcsname\relax
  \providecommand{\doi}[1]{doi: #1}\else
  \providecommand{\doi}{doi: \begingroup \urlstyle{rm}\Url}\fi

\bibitem[Amemiya(1985)]{amemiya1985advanced}
Takeshi Amemiya.
\newblock \emph{Advanced econometrics}.
\newblock Harvard university press, 1985.

\bibitem[Andersen(1992)]{andersen1992repeated}
Per~Kragh Andersen.
\newblock Repeated assessment of risk factors in survival analysis.
\newblock \emph{Statistical Methods in Medical Research}, 1\penalty0
  (3):\penalty0 297--315, 1992.

\bibitem[Andersen and Gill(1982)]{andersen1982cox}
Per~Kragh Andersen and Richard~David Gill.
\newblock Cox's regression model for counting processes: a large sample study.
\newblock \emph{The annals of statistics}, pages 1100--1120, 1982.

\bibitem[Cox(1972)]{cox1972regression}
David~R Cox.
\newblock Regression models and life-tables (with discussion).
\newblock In \emph{J. R. Statist. Soc.}, volume~B, pages 187--220. 1972.

\bibitem[Fan and Li(2001)]{fan2001variable}
Jianqing Fan and Runze Li.
\newblock Variable selection via nonconcave penalized likelihood and its oracle
  properties.
\newblock \emph{Journal of the American statistical Association}, 96\penalty0
  (456):\penalty0 1348--1360, 2001.

\bibitem[Fan and Li(2002)]{fan2002variable}
Jianqing Fan and Runze Li.
\newblock Variable selection for cox's proportional hazards model and frailty
  model.
\newblock \emph{Annals of Statistics}, pages 74--99, 2002.

\bibitem[Henderson et~al.(2000)Henderson, Diggle, and
  Dobson]{henderson2000joint}
Robin Henderson, Peter Diggle, and Angela Dobson.
\newblock Joint modelling of longitudinal measurements and event time data.
\newblock \emph{Biostatistics}, 1\penalty0 (4):\penalty0 465--480, 2000.

\bibitem[Hsieh et~al.(2006)Hsieh, Tseng, and Wang]{hsieh2006joint}
Fushing Hsieh, Yi-Kuan Tseng, and Jane-Ling Wang.
\newblock Joint modeling of survival and longitudinal data: likelihood approach
  revisited.
\newblock \emph{Biometrics}, 62\penalty0 (4):\penalty0 1037--1043, 2006.

\bibitem[Ibrahim et~al.(2010)Ibrahim, Chu, and Chen]{ibrahim2010basic}
Joseph~G Ibrahim, Haitao Chu, and Liddy~M Chen.
\newblock Basic concepts and methods for joint models of longitudinal and
  survival data.
\newblock \emph{Journal of Clinical Oncology}, 28\penalty0 (16):\penalty0
  2796--2801, 2010.

\bibitem[Kim et~al.(2013)Kim, Zeng, Li, and Spiegelman]{kim2013joint}
Sehee Kim, Donglin Zeng, Yi~Li, and Donna Spiegelman.
\newblock Joint modeling of longitudinal and cure-survival data.
\newblock \emph{Journal of statistical theory and practice}, 7\penalty0
  (2):\penalty0 324--344, 2013.

\bibitem[Lawrence~Gould et~al.(2015)Lawrence~Gould, Boye, Crowther, Ibrahim,
  Quartey, Micallef, and Bois]{lawrence2015joint}
A~Lawrence~Gould, Mark~Ernest Boye, Michael~J Crowther, Joseph~G Ibrahim,
  George Quartey, Sandrine Micallef, and Frederic~Y Bois.
\newblock Joint modeling of survival and longitudinal non-survival data:
  current methods and issues. report of the dia bayesian joint modeling working
  group.
\newblock \emph{Statistics in medicine}, 34\penalty0 (14):\penalty0 2181--2195,
  2015.

\bibitem[Rizopoulos(2011)]{rizopoulos2011dynamic}
Dimitris Rizopoulos.
\newblock Dynamic predictions and prospective accuracy in joint models for
  longitudinal and time-to-event data.
\newblock \emph{Biometrics}, 67\penalty0 (3):\penalty0 819--829, 2011.

\bibitem[Song et~al.(2002)Song, Davidian, and Tsiatis]{song2002semiparametric}
Xiao Song, Marie Davidian, and Anastasios~A Tsiatis.
\newblock A semiparametric likelihood approach to joint modeling of
  longitudinal and time-to-event data.
\newblock \emph{Biometrics}, 58\penalty0 (4):\penalty0 742--753, 2002.

\bibitem[Sousa(2011)]{sousa2011review}
In{\^e}s Sousa.
\newblock A review on joint modelling of longitudinal measurements and
  time-to-event.
\newblock \emph{Revstat Stat J}, 9:\penalty0 57--81, 2011.

\bibitem[Tibshirani(1996)]{tibshirani1996regression}
Robert Tibshirani.
\newblock Regression shrinkage and selection via the lasso.
\newblock \emph{Journal of the Royal Statistical Society. Series B
  (Methodological)}, pages 267--288, 1996.

\bibitem[Tsiatis and Davidian(2004)]{tsiatis2004joint}
Anastasios~A Tsiatis and Marie Davidian.
\newblock Joint modeling of longitudinal and time-to-event data: an overview.
\newblock \emph{Statistica Sinica}, pages 809--834, 2004.

\bibitem[Ye et~al.(2008)Ye, Lin, and Taylor]{ye2008semiparametric}
Wen Ye, Xihong Lin, and Jeremy~MG Taylor.
\newblock Semiparametric modeling of longitudinal measurements and
  time-to-event data--a two-stage regression calibration approach.
\newblock \emph{Biometrics}, 64\penalty0 (4):\penalty0 1238--1246, 2008.

\bibitem[Zeng and Lin(2007)]{zeng2007maximum}
D~Zeng and DY~Lin.
\newblock Maximum likelihood estimation in semiparametric regression models
  with censored data.
\newblock \emph{Journal of the Royal Statistical Society: Series B (Statistical
  Methodology)}, 69\penalty0 (4):\penalty0 507--564, 2007.

\bibitem[Zhang and Ringland(2017)]{XQZhang2017}
Xiaoqi Zhang and John Ringland.
\newblock A note on the joint probability density function of a stochastic
  growth process with random stopping time.
\newblock 2017.
\newblock URL \url{https://arxiv.org/abs/1701.04423}.

\bibitem[Zou(2006)]{zou2006adaptive}
Hui Zou.
\newblock The adaptive lasso and its oracle properties.
\newblock \emph{Journal of the American statistical association}, 101\penalty0
  (476):\penalty0 1418--1429, 2006.

\end{thebibliography}

\end{document}